\documentclass[12pt]{article}
\pdfoutput=1

\usepackage{epsf, amssymb}
\usepackage{cite}
\usepackage{epsfig}
\usepackage{graphicx}  
\usepackage{dcolumn}   
\usepackage{bm}        
\usepackage{amssymb}   
\usepackage{slashed}
\usepackage[utf8]{inputenc}
\usepackage[T1]{fontenc}
\usepackage{amsmath}
\usepackage{ae}
\usepackage{color}
\usepackage{graphicx}
\usepackage{bbm}

\newcommand{\arXiv}[1]{\href{http://www.arXiv.org/abs/#1}{#1}}
\usepackage[colorlinks=true, linkcolor=blue, bookmarks=true]{hyperref}

\def\be{\begin{equation}}
\def\ee{\end{equation}}
\newcommand  {\Rbar} {{\mbox{\rm$\mbox{I}\!\mbox{R}$}}}

\makeatletter
\renewcommand\section{\@startsection {section}{1}{\z@}%
                               {-3.5ex \@plus -1ex \@minus -.2ex}
                               {2.3ex \@plus.2ex}%
                               {\normalfont\large\bfseries}}
\renewcommand\subsection{\@startsection{subsection}{2}{\z@}%
                                 {-3.25ex\@plus -1ex \@minus -.2ex}%
                                 {1.5ex \@plus .2ex}%
                                 {\normalfont\bfseries}}
\makeatother

\begin{document}

\begin{titlepage}
 
\begin{center}
{\Large \bf Holographic Quenches and
\vspace{3mm} Fermionic Spectral Functions}

\vskip 10mm

{\large N.~Callebaut$^{a,b}$, B.~Craps$^{a,c}$, F.~Galli$^d$, D.~C.~Thompson$^a$, \\
\vspace{3mm}
J.~Vanhoof$^a$, J.~Zaanen$^e$, Hongbao~Zhang$^a$}
\vskip 7mm

$^a$ Theoretische Natuurkunde, Vrije Universiteit Brussel, and \\
\hspace*{0.15cm}  International Solvay Institutes, 
Pleinlaan 2, B-1050 Brussels, Belgium \\
$^b$ Ghent University, Department of Physics and Astronomy,\\ Krijgslaan 281-S9, 9000 Gent, Belgium \\
$^c$ Laboratoire de Physique Th\'eorique, Ecole Normale Sup\'erieure,\\ 24 rue Lhomond, F-75231 Paris Cedex 05, France \\
$^d$ Instituut voor Theoretische Fysica, KU Leuven,\\ 
Celestijnenlaan 200D, B-3001 Leuven, Belgium.\\
$^e$ Instituut-Lorentz for Theoretical Physics, Universiteit Leiden,\\ PO Box 9506, NL-2300 RA Leiden, The Netherlands

\vskip 10mm
 
{\small\noindent  {\tt ncalleba.Callebaut@ugent.be, Ben.Craps@vub.ac.be, federico@itf.fys.kuleuven.be, Daniel.Thompson@vub.ac.be, Joris.Vanhoof@vub.ac.be, Jan@lorentz.leidenuniv.nl, Hongbao.Zhang@vub.ac.be}}
 
\end{center}

\vskip 10mm

\begin{center}
{\bf ABSTRACT}
\vspace{3mm}
\end{center}
Using holographic methods we investigate the behaviour of fermionic spectral functions of strongly coupled $2+1$ dimensional field theories as both temperature and chemical potential are quenched.  
\vfill 
\end{titlepage}



\section{Introduction}

Arguably two of the most fundamental problems in quantum field theory, or equivalently quantum many body physics, are non-equilibrium unitary time evolution and finite density, respectively.  With the exception of integrable systems in 1+1D (see, for instance, \cite{Calabrese:2006rx}) there exists  no direct  field theoretical method that can deal with far from equilibrium dynamics in strongly interacting systems with  infinite   degrees of freedom.  Even at equilibrium, for strongly interacting fermionic systems at finite density, the fermion sign problem obscures the view of non-perturbative physics.

These problems, associated with understanding non-Fermi liquid physics, have been pushed to center stage
in condensed matter physics since the discovery of the strange metals in   high $T_c$ superconductors and related systems \cite{Anderson,VLSAR,VNS}. The study of non-equilibrium physics of quantum many body systems is at present rapidly evolving in the laboratory because of the new techniques becoming available in the field of cold atoms, high energy physics (heavy ion collisions)   and the ``ultrafast" experiments on condensed matter systems. 

The AdS/CFT correspondence is unique in its capacity to deal both with finite density and non-equilibrium in a mathematically controlled way, albeit within 
the limitations of the ``large $N$" limit and the restrictions on the non-equilibrium physics that can be addressed easily.  To illustrate this power, we present 
here an extreme example of a time-dependent ``experiment'' involving strongly interacting fermion matter. 

Consider a strongly interacting 
relativistic quantum critical state formed from fermions and other degrees of freedom in 2+1D at zero density and zero temperature. 
This can be pictured as graphene right at the  quantum phase transition to the Mott insulator \cite{Herbut}.  At large negative times its fermion
spectral functions, which can be measured by (inverse) photoemission, will be of the ``branch cut" form dictated by Lorentz and scale invariance: 
\begin{equation}
 A (\omega, k) \sim \text{Im} \left[(  \sqrt{k^2 - \omega^2})^{2\Delta-d}\right]  \ .
\end{equation}
Here,  $d$ is the spacetime dimension and throughout this work we will consider  $d=3$.  We are typically interested in the regime  $2\Delta - d < 0$, noticing that the scaling dimension is bounded from below by the unitarity limit for the fermionic CFT operators, $\Delta = (d-1)/2$. 

Subsequently, this system is prepared in a highly excited state by a sudden ``quench" at a given instant, but now at some finite {\em charge density} which is uniform in space. The main purpose of this paper is to compute time-dependent spectral properties of this system using holography. Specifically, we compute its time-dependent spectral function and show that it suggests the gradual build-up of a Fermi surface. We also highlight some caveats, commenting in particular on the relation with quantities potentially measurable using time-resolved ARPES experiments. 


\section{Time-dependent fermion spectral functions}

In equilibrium, the spectral function of an operator  is defined as ($-2$ times) the imaginary part of its corresponding retarded Green's function in frequency space. However, out of equilibrium such a definition is ambiguous;  the retarded Green's function $G_R(\vec{x}_1, t_1; \vec{x}_2, t_2)$ may no longer depend on just the relative time interval $  t =  t_2  - t_1$ but on $t_1$ and $t_2$ separately and  a conventional Fourier transform to frequency space can not be taken. 
In \cite{Balasubramanian:2012tu}, a  generalised notion of a spectral function was proposed and computed in a far-from-equilibrium AdS/CFT context
(see \cite{CaronHuot:2011dr,Chesler:2011ds} and \cite{Banerjee:2012uq,Mukhopadhyay:2012hv,Steineder:2013ana} for studies of this and related quantities far from equilibrium and closer to equilibrium, respectively).  
Introducing both the relative time $t$ and the average time, $T=(t_1+t_2)/2$, 
   we consider the Fourier transform of the  Green's  function with respect to  $t$ whilst keeping $T$ fixed, 
   \begin{equation}\label{eq:Wignertrans}
 G(T, \omega, k) = \int dt~e^{i \omega t} G(T,t, k)  \ .
  \end{equation}
We assume  that  homogeneity   is preserved such that we can Fourier transform to spatial momenta with no change. Then a notion of a time-dependent spectral function is defined by 
  \def\Im{{\rm{ Im}}}
  \begin{equation}\label{eq:SpectralDef}
  A(T,\omega, k) = - 2 \, \Im  \, G_R(T, \omega, k) \ . 
  \end{equation}
In passing we note that this recipe for taking a Fourier transform has a rather well established analogue as the Wigner distribution \cite{Wigner:1932eb} used in phase space approaches to Quantum Mechanics.

Suppose that  a system starts in equilibrium, and then undergoes an instantaneous quench at time $t_0=0$, eventually relaxing to find a new equilibrium. 
At $T=-\infty$ the spectral function \eqref{eq:SpectralDef} will match the conventional equilibrium spectral function of the initial configuration and at  $T=+\infty$ it will match that of the system at its new equilibrium.  
For intermediate values of average time,  \eqref{eq:SpectralDef} can generically display rather wild oscillations and indeed need not remain positive for positive frequencies (see, for instance, \cite{Balasubramanian:2012tu}). This feature is similar to the ``negative probabilities'' displayed by the Wigner distribution and may be countered by coarse-graining with a Gaussian filter.  
 
One consequence of the inherent non-locality of the definition \eqref{eq:SpectralDef} is that for $T<0$ the spectral function is already influenced by the effects of the quench; this does not represent non-causality, it is simply because for a large enough relative time, the interval centered around $T$ will necessarily extend through the time of the quench.   
One could choose an alternate definition removing this feature, for instance by considering the retarded Green's function as a function of final time $t_2$  and relative time $t=t_2-t_1$, and Fourier transforming with respect to the relative time $t$ at fixed final time. (Note that the retarded Green's function vanishes for $t<0$.)


\section{Holographic setup}  

How can a charged quench be described in AdS/CFT? In AdS/CFT the field theoretical problem in $d$ dimensions is dualized in an equivalent gravitational problem living in a $d+1$ dimensional bulk, described by classical gravity in the large $N$ limit of  the  gauge theory. When the bulk is characterised by an Anti-de-Sitter (AdS) geometry it describes the vacuum of a conformal field theory on
the boundary at zero temperature and  density. To study the non-equilibrium  physics associated with an instantaneous quench in the zero density theory, a standard simple approach (see, for instance, \cite{Bhattacharyya:2009uu,Balasubramanian:2010ce,Wu:2012rib,Hubeny:2007xt,AbajoArrastia:2010yt,Albash:2010mv,Liu:2013iza, Balasubramanian:2011at,Allais:2011ys,Callan:2012ip,Hubeny:2013hz})  is to inject at the AdS boundary a shell of light-like ``dust" (or other matter that effectively behaves like dust) in  AdS$_{d+1}$. This corresponds in the field theory with the sudden creation of an excited state whose time evolution is subsequently determined by the equivalent gravitational evolution in the bulk.   A time-dependent metric describing such a process is available in closed form; it is the AdS version of a metric derived by Vaidya in the 1930's.   After some time this in-falling shell of dust will form a black brane, which in turn encodes a thermal equilibrium state in the field theory: this is the holographic description of the anticipated thermalization of the field theory in the long time limit. 

\begin{figure}[ht] 
\centering 
\includegraphics[width=0.28 \textwidth]{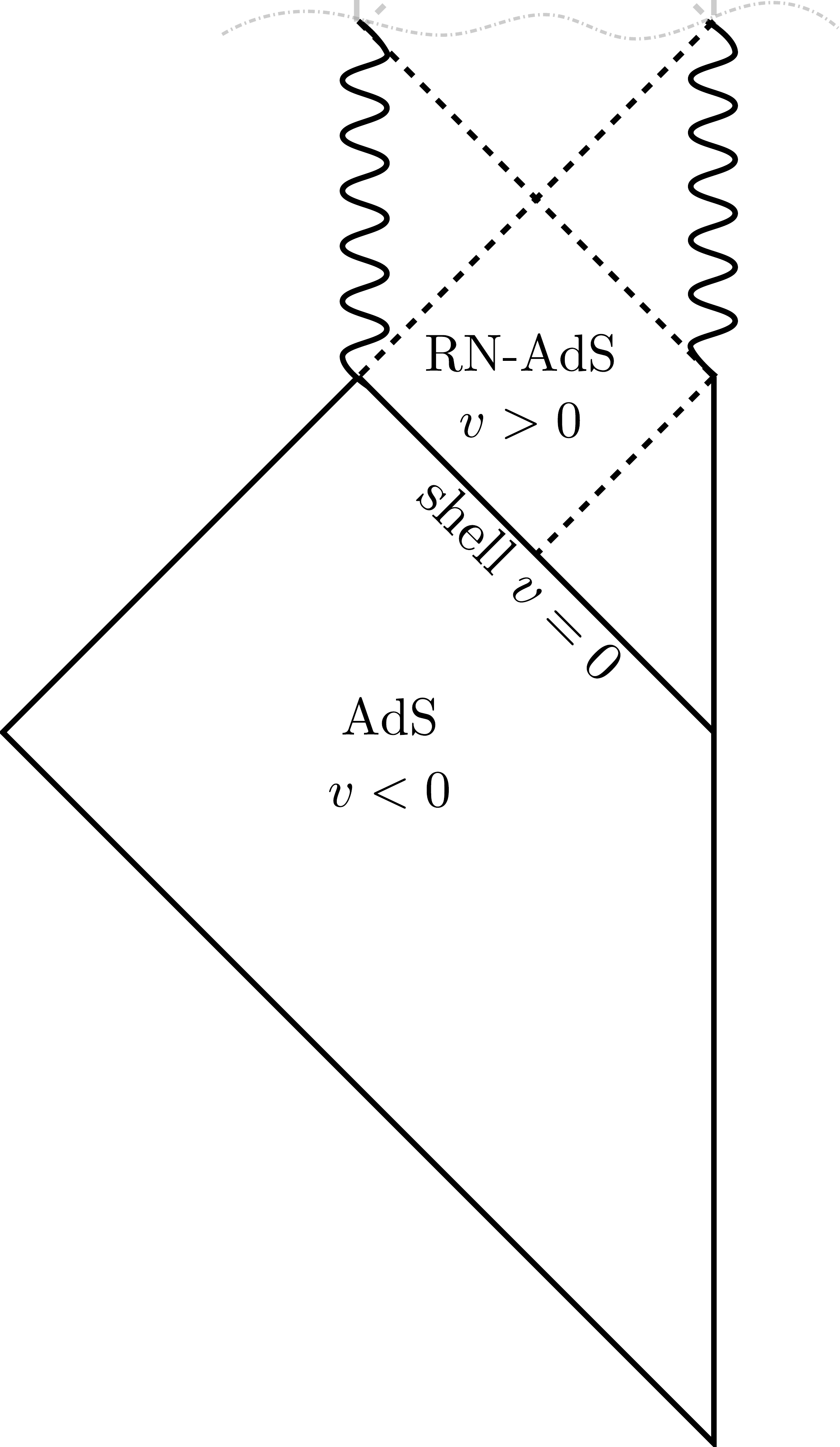}
\caption{ Penrose diagram for Vaidya AdS Reissner-Nordstr{\"o}m (RN) spacetime. }\label{fig:Penrose}
\end{figure}

This in-falling shell paradigm of holographic thermalisation was recently extended to Einstein-Maxwell gravity, where one can consider the  injection of a shell of dust which is {\em electrically charged} \cite{Galante:2012pv,Caceres:2012em}.  Fig.~\ref{fig:Penrose} illustrates the Penrose diagram for the corresponding Reissner-Nordstr\"om-AdS-Vaidya spacetime.
Although the (scale-dependent) equilibration time becomes dependent on whether the chemical potential is large or small 
compared to temperature, the gross evolution in the bulk is similar as in the zero density case: after some time a charged Reissner-Nordstr{\"o}m (RN) black brane will form, describing a holographic strange metal at finite temperature.

It has been known for a while that, for sufficiently relevant scaling dimensions of the fermion operators of the zero density CFT,  Fermi liquid like 
quasiparticles form  at finite density \cite{Liu:2009dm,Cubrovic:2009ye,Faulkner:2009wj}, characterised by a propagator  
 \be\label{fermion2pt}
G(\omega,k)=\frac{1}{\omega-v_F(k-k_F)+\Sigma(\omega,k)}
\ee
with self energy $\Sigma \sim \omega^{2\nu_{k=k_F}}$ where $\nu_k$ is associated with the scaling dimension in the AdS$_2\times \Rbar^2$
near horizon geometry. We deliberately choose parameters such that $2 \nu_{k_F} > 1$, so that sharp quasiparticles are formed in the finite density 
system.

It is however by now well understood that, when backreaction from bulk fermions is taken into account, the RN black hole is quantum mechanically unstable to the formation of an electron star \cite{Hartnoll:2009ns,Hartnoll:2010gu,Cubrovic:2010bf,Cubrovic:2011xm,Allais:2013lha} (see \cite{Hartnoll:2011fn} for a review). This also corresponds to a thermodynamic instability, in the sense that for fixed temperature and chemical potential, the electron star solution has lower free energy than the RN black brane. While it would be interesting to study this effect in the present fixed-energy (as opposed to fixed-temperature) context, in the present paper we choose to work in the limit in which we treat the bulk classically (which can be thought of as ignoring perturbative and non-perturbative $1/N$ corrections).
 
In our $d=3$ case, the metric and the bulk spacetime gauge fields are  given  in Eddington-Finkelstein coordinates    by 
\begin{equation}
\begin{aligned}
&ds^2 = \frac{1}{z^2} \left(-f(v,z) dv^2 -  2 dv dz  + d\vec{x}^2 \right) \ , \\
&A = g(v,z) dv  \, , 
\end{aligned}
\end{equation}
where
\begin{equation}
\begin{aligned}
f(v,z) &=1+  \Theta(v)\left( - ( 1 + Q^{2}) z^3 +  Q^2 z^4 \right)  \ , \\
\quad g(v,z) &= \Theta(v) \mu (1- z)  \ . 
\end{aligned}
\end{equation}
At the  spacetime boundary $z=0$, the bulk lightcone time $v$ coincides with the field theory time.
As in \cite{Liu:2009dm} we work with dimensionless quantities obtained by rescaling such that the horizon is fixed at $z=1$. The temperature of the final black hole produced is  $\mathcal T=\frac{1}{4\pi} \left(3 - Q^2 \right)$  and the parameter $\mu= g_F Q$ can be identified with the chemical potential of the theory where $g_F$ is the $U(1)$ coupling which we shall set to unity.

To determine fermionic spectral functions, we use the same strategy that was employed in \cite{Balasubramanian:2012tu} for scalar spectral functions. The retarded propagator of an operator is determined by computing its expectation value following a perturbation by a delta-function source \cite{Iqbal:2008by}. The expectation value and the source are identified with certain coefficients in a near-boundary expansion of the dual bulk field. The bulk field is obtained by solving its equation of motion subject to boundary conditions determined by the delta-function source and initial conditions that set the field to zero outside the future lightcone of the source.  

To implement this procedure for a fermionic operator of conformal dimension $\Delta$ and $U(1)$ charge $q$, we consider the Dirac equation for a bulk fermionic field of mass $m$, 
\begin{equation}
(\slashed{D} - m)\Psi \equiv \gamma^{\mu}\partial_{\mu}\Psi  + \frac{1}{4} \gamma^{\mu} \omega_{\mu AB}\Gamma^{AB}\Psi - i q \gamma^{\mu}A_{\mu} \Psi  - m \Psi = 0  \ , 
\end{equation} 
in which $\gamma^\mu$ are the curved space gamma matrices of the $AdS$ spacetime, $\omega_{\mu AB}$ the spin connection, $\Gamma^A$ the constant bulk gamma matrices in the local tangent frame and  $\Gamma^{AB} = \frac{1}{2} [\Gamma^A,\Gamma^B]$ the antisymmetric generators of Lorentz transformations. 

The conformal dimension is related to the mass of the bulk field by $(\Delta-3/2)^{2}=m^{2}$ or $\Delta_{\pm}=3/2\pm m$. One should keep in mind that the unitarity bound on fermionic CFT operators is $\Delta>(d-1)/2=1$. This implies that in the mass range $m<-1/2$ only the choice $\Delta=\Delta_{-}$ is allowed and in the range $1/2<m$ only $\Delta=\Delta_{+}$. In the mass range $-1/2<m<1/2$, both $\Delta=\Delta_{-}$ and $\Delta=\Delta_{+}$ are allowed. These two possibilities are related by switching the roles of the source and the expectation value in the identification with coefficients of the bulk field. Throughout this paper, we are assuming that the conformal dimension of the operator is 
\begin{equation}
\Delta=\Delta_{+}=\frac32+m.
\end{equation} 
We will consider operators of  conformal dimension $\Delta \leq 3/2$ (with the main focus of the analysis on the case $m=0$).  Spectral functions of fermionic operators in this conformal dimension range have been studied at equilibrium in the RN geometry  in \cite{Liu:2009dm,Cubrovic:2009ye,Faulkner:2009wj}.    

To proceed, we will choose the spatial momentum to be aligned in the  $x^{1}$ direction and define
\begin{equation} \label{eq:spinorresc}
\Psi = z^{m+ \frac{1}{2}} e^{i k x^{1}  }   \left(  \begin{array}{c} y_{+} (v,z) \\  z_{+} (v,z) \\ y_{-} (v,z) \\ z_{-} (v,z)\end{array} \right) \ . 
\end{equation}
With a  judicious choice of gamma matrices, 
\begin{equation}
\Gamma^{v}=\left( \begin{array}{cc}  0& i \sigma_{2} \\ i\sigma_{2} & 0  \end{array}\right) \ , \quad
\Gamma^{z}=\left( \begin{array}{cc}  \mathbb{I}_{2} & 0 \\ 0 & -\mathbb{I}_{2} \end{array}\right) \ , \quad
\Gamma^{1}=\left( \begin{array}{cc}  0&  \sigma_{1} \\   \sigma_{1} & 0  \end{array}\right) \ , \quad
\Gamma^{2}=\left( \begin{array}{cc}  0&  \sigma_{3} \\   \sigma_{3} & 0  \end{array}\right) \ , \quad
\end{equation}
and frame fields,
 \begin{equation}
 e^{v} = \frac{1}{z}\left(d z +\frac{1}{2} ( f(v,z)+1) dv \right)\ , \quad e^{z} = \frac{1}{z}\left(d z +\frac{1}{2} (f(v,z)-1) dv \right)\ , \quad    e^{x^{i}  } = \frac{1}{z}dx^{i}\ ,
  \end{equation}
one finds the equations of motion for $y_{+}$ and $z_{-}$ decouple from those of $y_{-}$ and $z_{+}$  \cite{Liu:2009dm}.  It suits our purpose to define the combinations  $\alpha = y_+ + z_-$ and $\beta= y_+- z_-$,
in terms of which the equations of motion become 
\begin{equation}
\begin{aligned} 
0&=  (-1 + m) \alpha  - (m + i k z) \beta + z \partial_{z}\alpha  \ ,    \\
0&=  -2 (m - i k z) \alpha + \left( z \partial_{z}f +  2 f (m-1 )+4 i q z g \right) \beta  + 2 z f \partial_{z}\beta - 4 z \partial_{v}\beta \ .  \label{eq:eqms}
\end{aligned}
\end{equation}  
Similar equations for $y_{-}$ and $z_{+}$ can be obtained by substituting $k \leftrightarrow - k$ in these.

Since we are working in a mixed representation, in which only spatial momenta are Fourier transformed at first, the prescription for finding the  bulk-to-boundary retarded Green's  function is to simply   consider the causal response to a delta-function source on the boundary. Then, using the standard holographic dictionary identifying the source and vev of the boundary CFT operator with the asymptotic behaviour of the bulk field \cite{Henningson:1998cd,Iqbal:2009fd}, one can extract the boundary retarded Green's function $G_R$. Whilst $G_R$ is a $2\times2$ matrix, in this basis of gamma matrices it is diagonalised.  

Specifically, $z_{-}$ and $y_{+}$  have the asymptotic expansion
\be
\begin{aligned}
z_{-} &\approx z^{1-2m} (A + O(z)) + z^{2} (B + O(z)) \, ,\\
y_{+} &\approx z^{2-2m} (C + O(z)) + z ~(D + O(z))\, ,
\end{aligned}
\ee
where $A$ is identified as the source and $D$ is identified as the expectation value. $C$ and $B$ are local functions of $A$ and $D$ (and of their derivatives), respectively, as determined by the asymptotic Dirac equation.\footnote{See for instance \cite{Liu:2009dm}. Notice however the different rescaling and chirality conventions.}

To invoke a delta-function source at a time $t_1$ on the boundary one thus enforces 
\begin{equation}
\lim_{z\to0} z^{2m -1} z_{-} (v,z)=\lim_{z\to 0} z^{2m-1}\frac{\alpha(v,z) -\beta(v,z)}{2} = \delta(v-t_1) \ ,
\end{equation}
and to ensure that the propagator is causal (i.e., the retarded propagator) we impose that 
\begin{equation}
\beta(v<t_1, z ) = 0 \ . 
\end{equation}
By virtue of the Dirac equation, this is sufficient to ensure that also $\alpha(v<t_1,z)= 0$. With these conditions one can then determine the appropriate solution to \eqref{eq:eqms}. 

For $t_2 > t_1$ the source has no support and the expectation value of the boundary operator can be easily extracted. The upper component of $G_R$ can  be calculated as  
\begin{equation} 
G_{11}(t_2, t_1)  = -i \lim_{z\to0} z^{-1} y_{+}(v_2,z) = -i \lim_{z\to0} z^{-1} \frac{\alpha(v_2,z) +\beta(v_2,z)}{2}  \ ,  
\end{equation} 
with a similar  definition for the lower component $G_{22}$ coming from the analogous equations for  $y_{-}$ and $z_{+}$. 


\section{Numerical solution strategy}

To solve the equations of motions  \eqref{eq:eqms}  we can distinguish three cases: a) when the relative time interval  $t=t_2-t_1$ occurs entirely before the quench; b) when it extends across the quench and c) when it occurs entirely after the quench.  

In case a) one needs only to consider the pure AdS region of the Vaidya spacetime for which exact analytic expressions are known for the bulk-to-boundary retarded Green's function.  

 In case b)  the same analytic expressions may be used for the AdS portion of the spacetime and in particular on the shell of null dust at $v=0$.  We then   employ numerical methods to propagate these analytic expressions forwards from the shell at $v=0$ to fill out the remainder of the spacetime outside the horizon. 
In particular, knowing $\beta$ on a fixed $v$ slice, the first equation in \eqref{eq:eqms} with suitable boundary conditions can be integrated to obtain the bulk profile  for $\alpha$.  Given the bulk profiles for $\alpha$  and $\beta$ on a fixed $v$ slice the second equation in \eqref{eq:eqms} is used to evolve $\beta$ to a subsequent time. 
To perform the integration in the radial, $z$, direction we employ a  Chebyshev pseudospectral method, combined with an explicit $4^{th}$ order Runge-Kutta method in the $v$ direction.   For numerical purposes, it is also necessary to impose a cut-off at late $v_{cut}$ in our numerical integration.  To generate the plots in the next section we took  $N=27$ Chebyshev points, a Runge-Kutta time step $\delta v= 0.05$ and a cut-off in the Runge-Kutta integration $v_{cut}=60$.   Details of the convergence and accuracy of these methods are provided in appendix A where we show that for the chosen values of parameters the results for the spectral function are convergent to within the resolution of plots shown.
 
In case c), although the relevant geometry is the AdS-RN black hole, one does not have access to analytic expressions for the retarded propagator in the mixed representation $G_R(t, k)$.  We found a sensible approach in this case to  first calculate $G_R(\omega, k)$ in Fourier space  as in  \cite{Liu:2009dm}  and to then perform an inverse Fourier transform to find the desired Green's function in the mixed representation.     

The next step to obtain our time-dependent spectral function is to perform the Wigner transformation to Fourier space as described  in \eqref{eq:Wignertrans}. Depending on the conformal dimension of the operator,   $G_R(T,t, k)$ can exhibit power law divergences as $t\rightarrow 0$, which need to be appropriately regulated.  To address  this problem one can take  
 the ratio of $G_R(T,t, k)$ with another analytically known and appropriately regulated function ${\cal A}(t)$  such that the ratio is finite as $t\rightarrow 0$.  If the Fourier transform of  ${\cal A}(t)$ is also known analytically, then a simple application of the convolution theorem can be used to establish  $G_R(T, \omega, k)$.   In this work, Fourier transformations are implemented numerically using discrete Fourier routines.  To avoid artefacts associated to the late time cut-off $v_{cut}$ of our numerical evolution, e.g.\ Wilbraham-Gibbs or ringing phenomena,  we include a Lanczos $\sigma$ factor in our Fourier transformation \cite{Lanczos}. 


\section{Results}

\begin{figure}[ht]
 \begin{center}
\includegraphics[width=0.32 \textwidth]{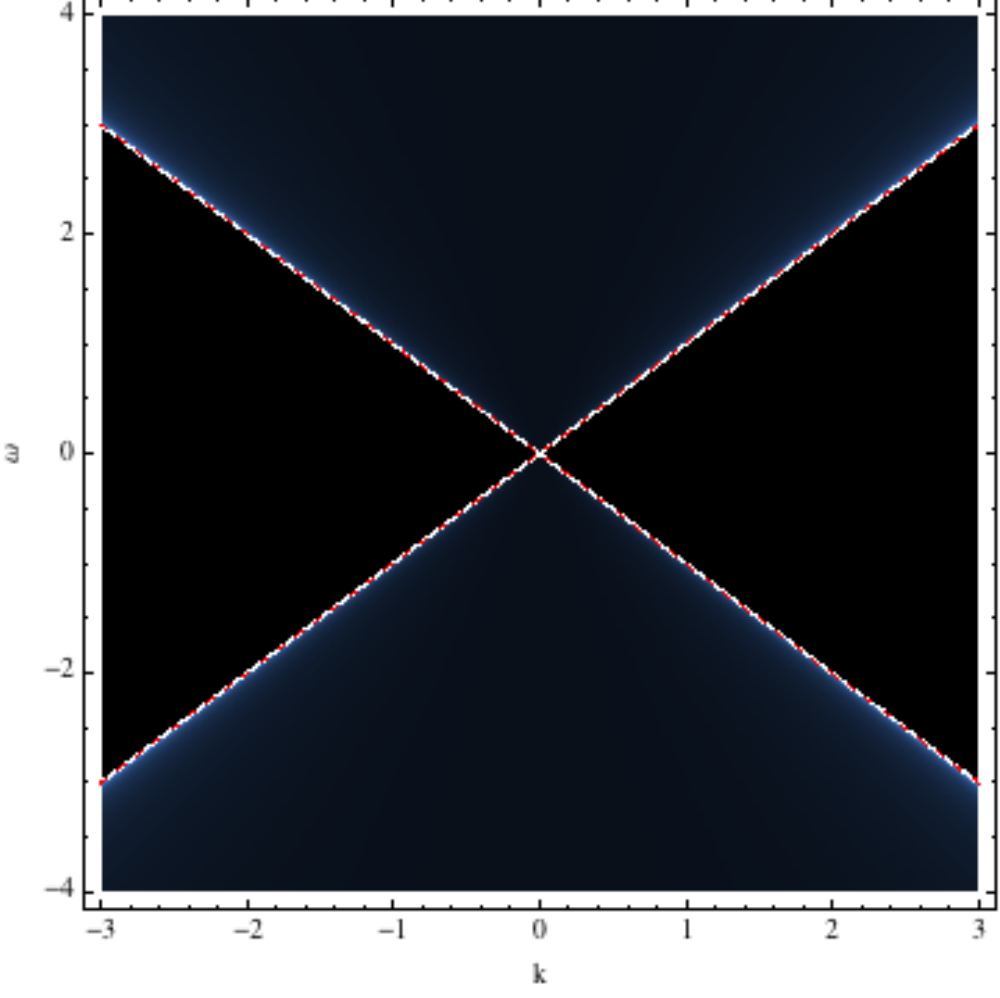} 
 \includegraphics[width= 0.32 \textwidth]{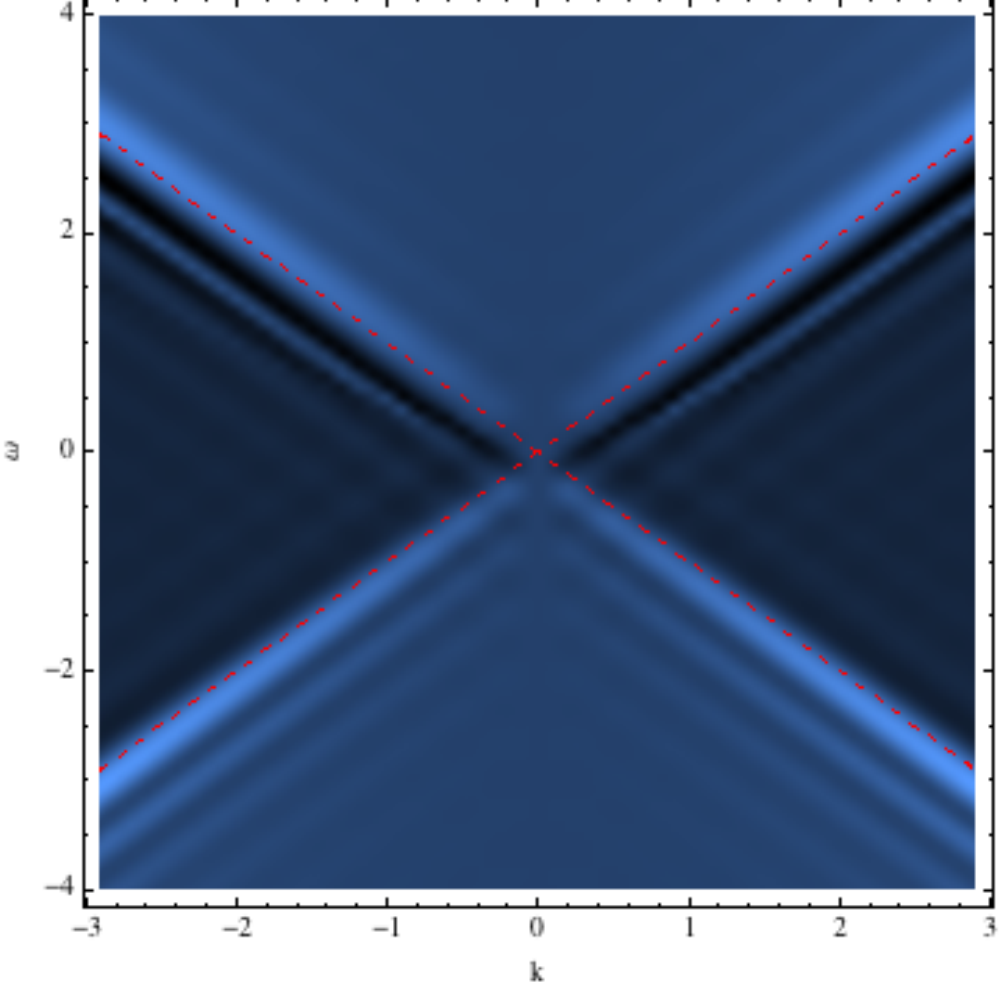}  
\includegraphics[width=0.32 \textwidth]{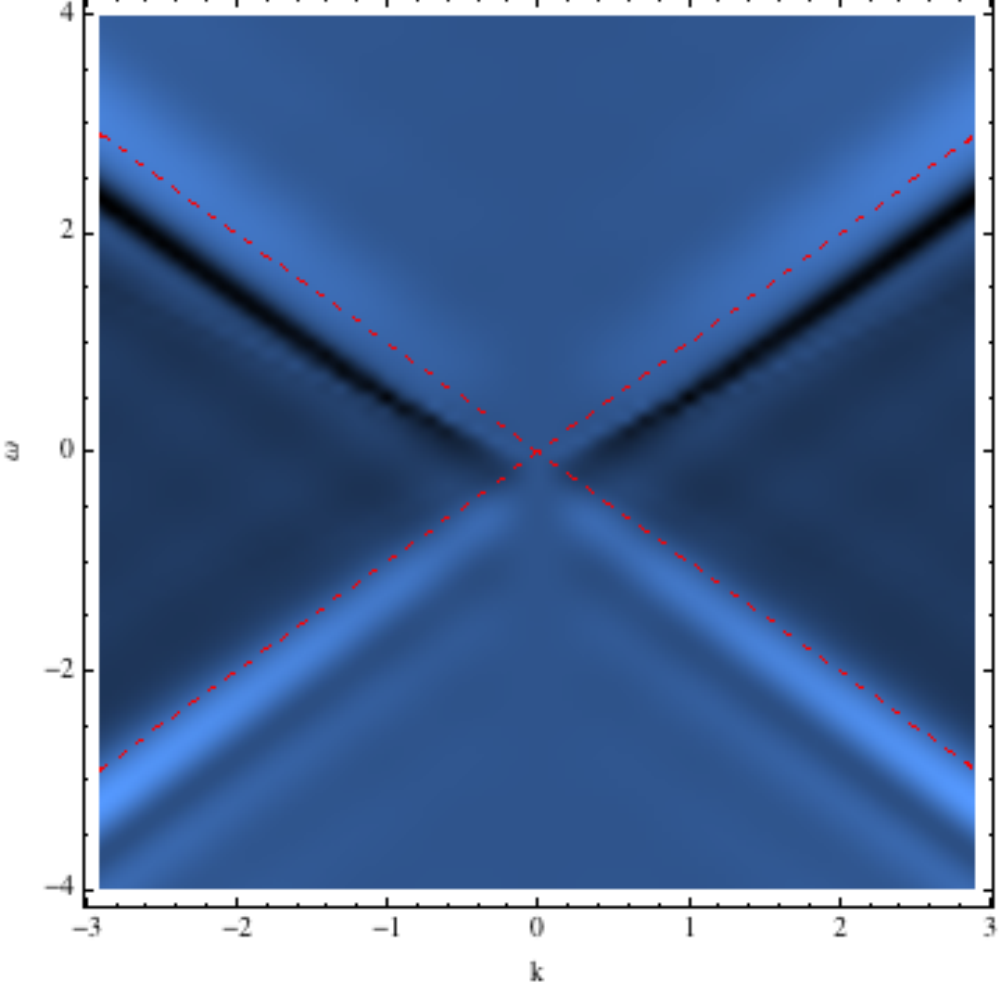}  \\
 \includegraphics[width=0.32 \textwidth]{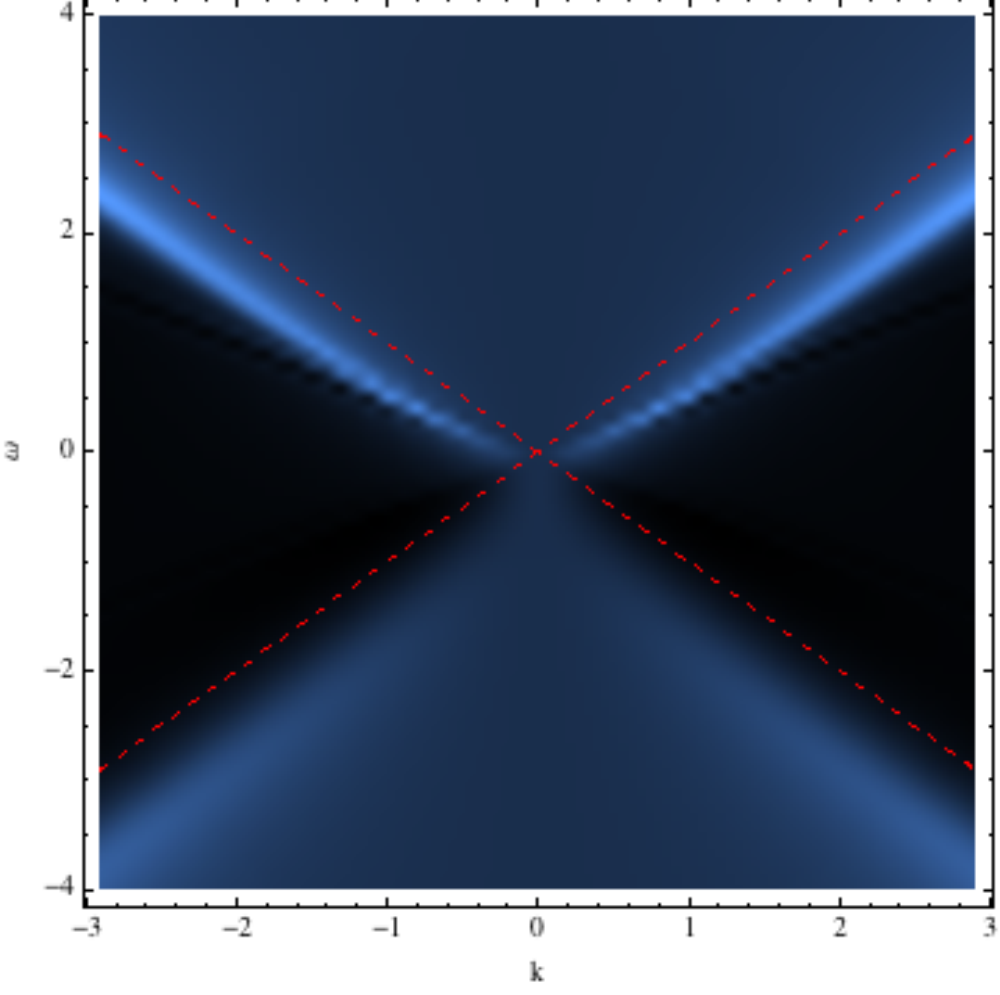} 
 \includegraphics[width=0.32\textwidth]{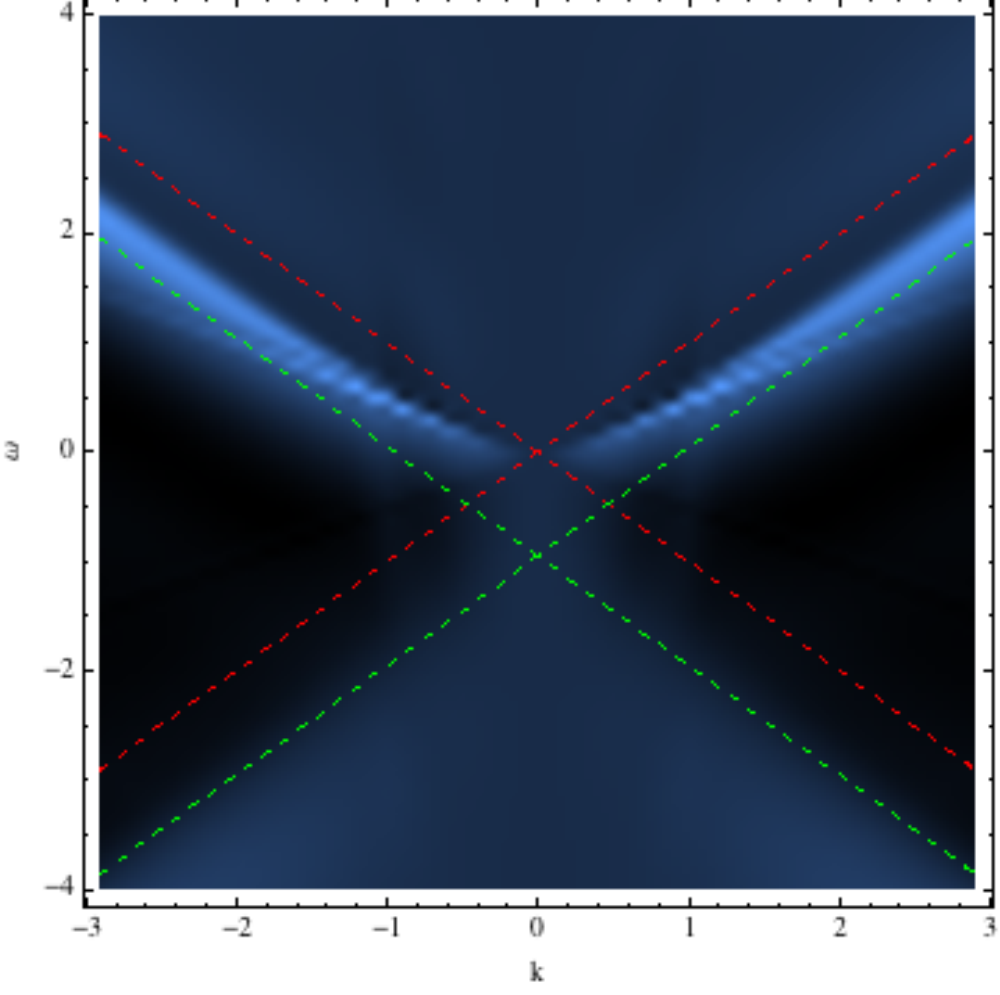}
\includegraphics[width=0.32\textwidth]{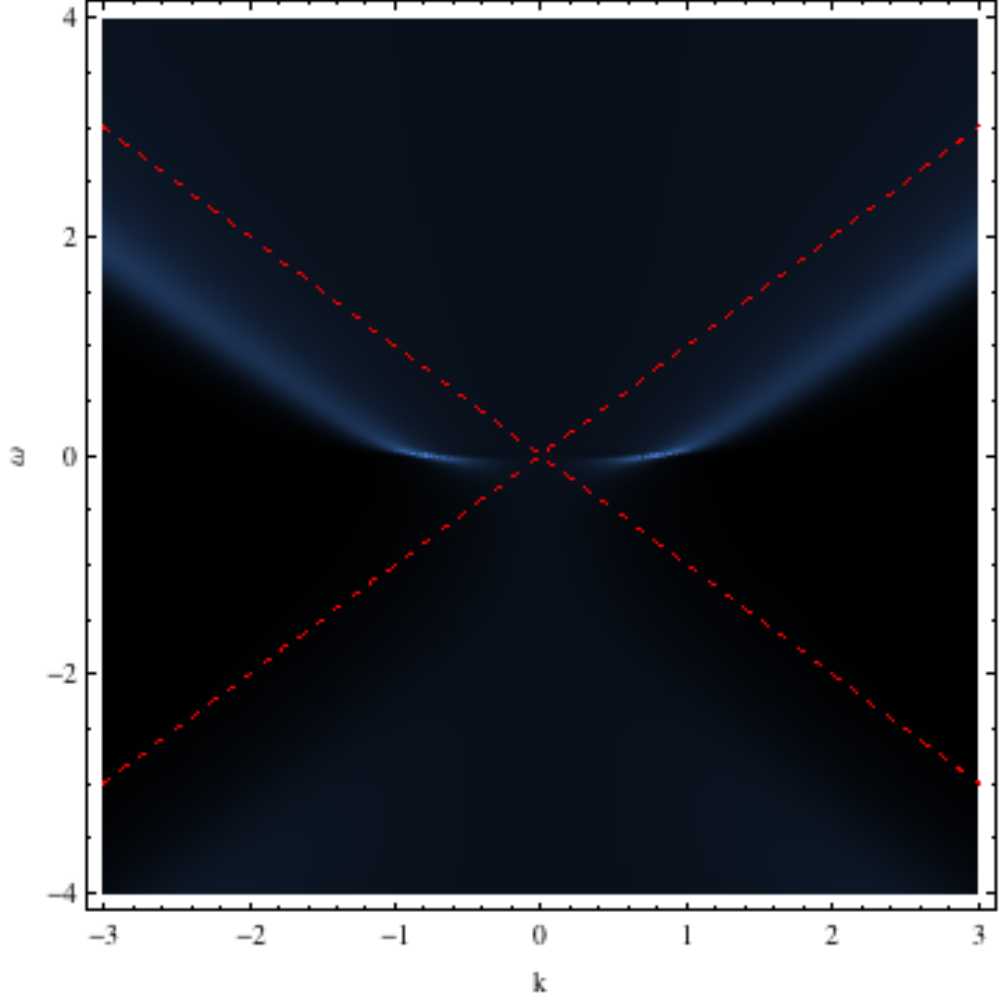}
  \end{center}
\caption{False colour density plots  for spectral sum $A= -2 \left[ \Im(G_{11})+ \Im(G_{22})\right]$  for bulk field of mass $m=0$ when quenched to $Q=1.7$,   at   (from  top to bottom) $T=\{-\infty,
-5,-2.5,  0,   1.5,+\infty
\}$. For reference we plot, in red dashed lines, lightcones centered around $\omega =0$.} 
  \label{fig:DensityPlots}
\end{figure}

Fig.~\ref{fig:DensityPlots} shows the spectral sum (the trace of the spectral function matrix),
\be
A(T, \omega, k)=-2\left[ \Im(G_{11}(T, \omega, k))+ \Im(G_{22}(T, \omega, k))\right] \ , 
\ee
in the $\{\omega, k\}$ plane for a fermionic operator of conformal dimension $\Delta = 3/2$ (dual to a bulk field of mass $m=0$)  when we quench the system to a near extremal final state:  the chemical potential is quenched from $\mu =0 $ to $\mu = 1.7$ and the temperature is kept close to zero (recall at zero temperature $Q=\sqrt{3} \approx 1.732$).
A small non-zero temperature has the advantage of stabilising the numerics, while being closer to what one would obtain in a real experimental situation. 

\begin{figure}[h]
  \begin{center}  
      \includegraphics[width= 0.45\textwidth]{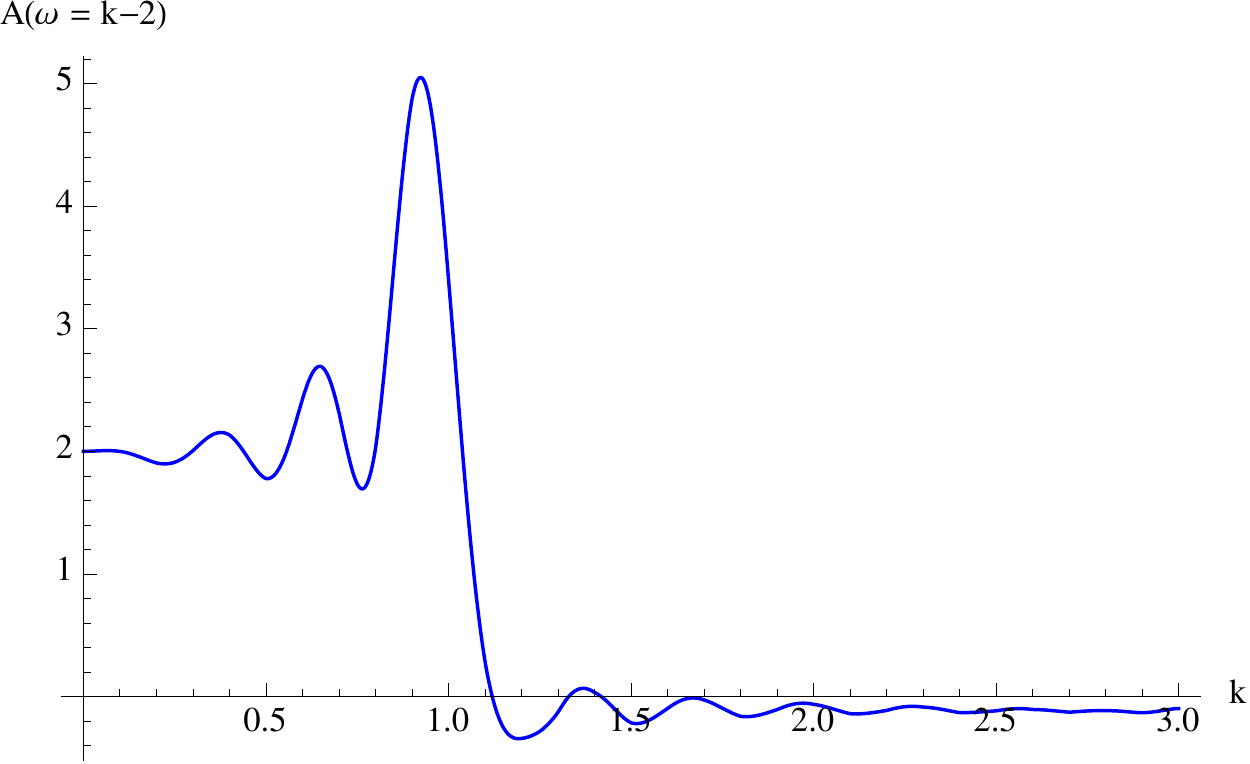} \qquad
      \includegraphics[width= 0.45\textwidth]{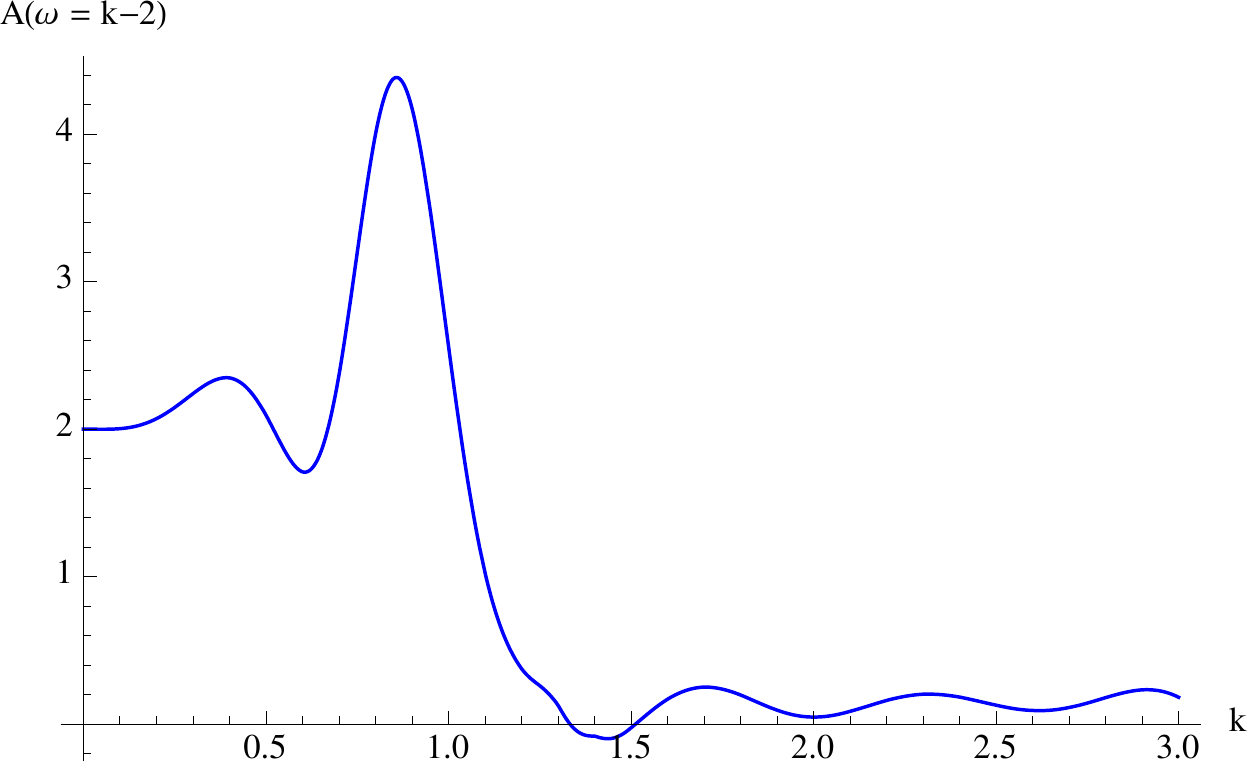} 
 \end{center}
  \caption{Detail of section of  spectral sum along  $\omega = k-2$ at average times   $T=-5$ (left) and $T = -2.5$ (right) highlighting an approximate halving of the frequency of oscillations.}
  \label{fig:SpectralSection}
\end{figure}

The top central panel of Fig.~\ref{fig:DensityPlots} shows the result at an early average time $T=-5$ and we see that whilst the density   is roughly symmetric around the light cone centered at $\omega=0$ (indicated in the figure by the dashed red lines), a number of oscillations have already begun to influence the profile. In the top right panel, at $T=-2.5$, the period of the oscillations has roughly doubled, as can be seen more clearly in  Fig.~\ref{fig:SpectralSection}.

In the  bottom left panel, at $T=0$, the oscillations have essentially disappeared, and the spectral function has developed a clear asymmetry between positive and negative frequency branches.  In the bottom central panel, at $T=1.5$, it is now clear that the asymptotic behaviour of the peak of the spectral function has shifted down still further, characteristic of a system at finite density.

Taking Fig.~\ref{fig:DensityPlots} at face value, after a transient regime at short times, characterised by oscillations  of the ``critical Dirac cones", it appears that at longer times a Fermi energy and Fermi surface develop. Observing that the asymptotic ``Dirac cones'' gradually move down, it is tempting to define a notion of time-dependent effective Fermi energy, or equivalently time-dependent effective chemical potential, $\mu_{eff}(T)$, which measures how much the asymptotic cones have moved down. (This is indicated by the green dashed lightcone in the bottom central panel of Fig.\ \ref{fig:DensityPlots}.)\footnote{At a practical level we numerically calculate the location of the maximum value $\omega_{\star}$ of the spectral sum at a fixed large $k$ and use this to define  $\mu_{eff}(T)= k-\omega_{\star}$. However, since the peaks of the spectral functions have a finite width and a profile that depends on both $Q$ and $\Delta$, in what follows to make true comparisons we always take the ratio of this quantity with the same quantity calculated in the final equilibrium state.}

\begin{figure}[h]
  \begin{center}
    \includegraphics[width=0.5\textwidth]{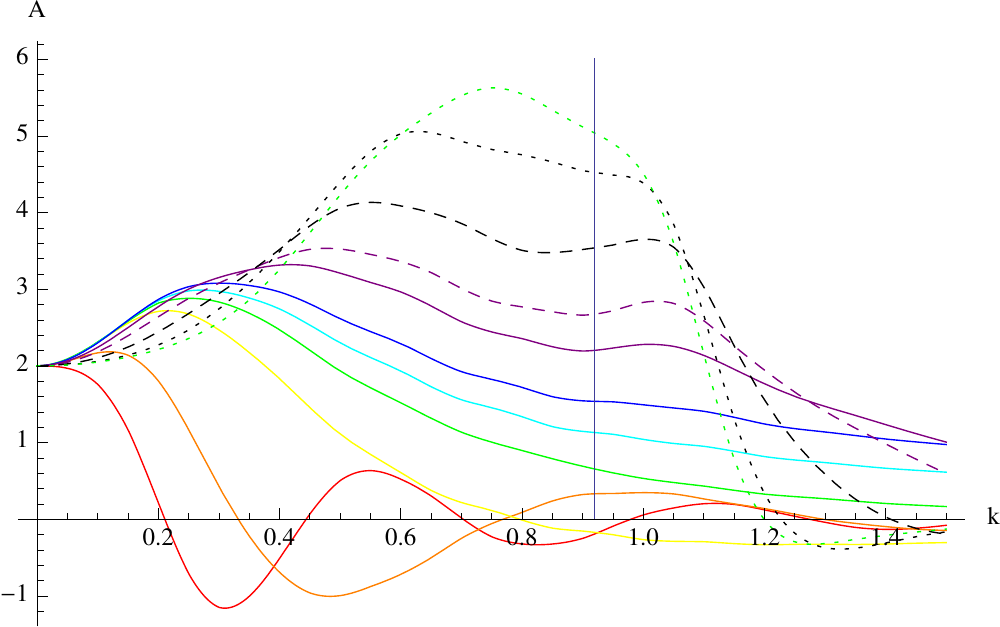}
       \end{center}
\caption{Plots of spectral sum at $\omega = -10^{-3}$ for field of mass $m=0$ when quenched to $Q=1.7$ for $T= \{-5,-2.5,-0.5,0,0.25, 0.5, 1.5,2.5,5, 10,15 \}$. The higher $T$, the higher the maximal value.  Also shown is a line at $k_F \approx 0.92$, the value of the Fermi momenta established in \cite{Liu:2009dm}. }   \label{fig:omega0Plots}
\end{figure}

Fig.~\ref{fig:omega0Plots} 
shows the accumulation of a peak at zero frequency for a number of average times.  We see for very early times (the curves with lowest peaks in Fig.~\ref{fig:omega0Plots}) that  the spectral function attains its maximum very close to zero momenta, as would be expected at zero density,  and exhibits clear oscillations including negative spectral weight.  As the average time evolves one finds that these oscillations die out, the location of the maximum at zero frequency migrates to larger values of $k$ and the value at the maximum increases.  The edge at larger $k$ sharpens and a clear peak develops with the location of the peak moving to where ultimately a sharp spike at $k=k_F \approx 0.92$ \cite{Liu:2009dm} will occur.

\begin{figure}[h]
  \begin{center}
    \includegraphics[width=0.5\textwidth]{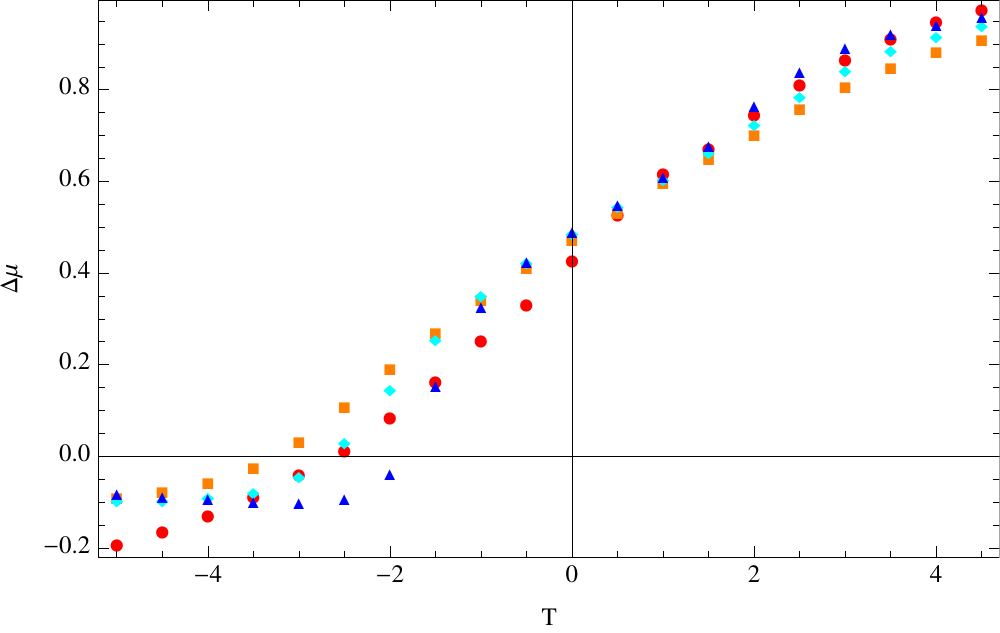}
       \end{center}
\caption{Plots of the relative effective chemical potential $\Delta \mu(T) = \mu_{eff}(T)/\mu_{eff}(+\infty)$ as a function of average time, $T$, for $m=0$ quenches with $Q = 0.5$ (red circles), $1$ (orange squares), $1.25$ (cyan diamonds), $1.7$  (blue triangles).  }   \label{fig:chempot}
\end{figure}

\begin{figure}[h]
  \begin{center}
    \includegraphics[width=0.5\textwidth]{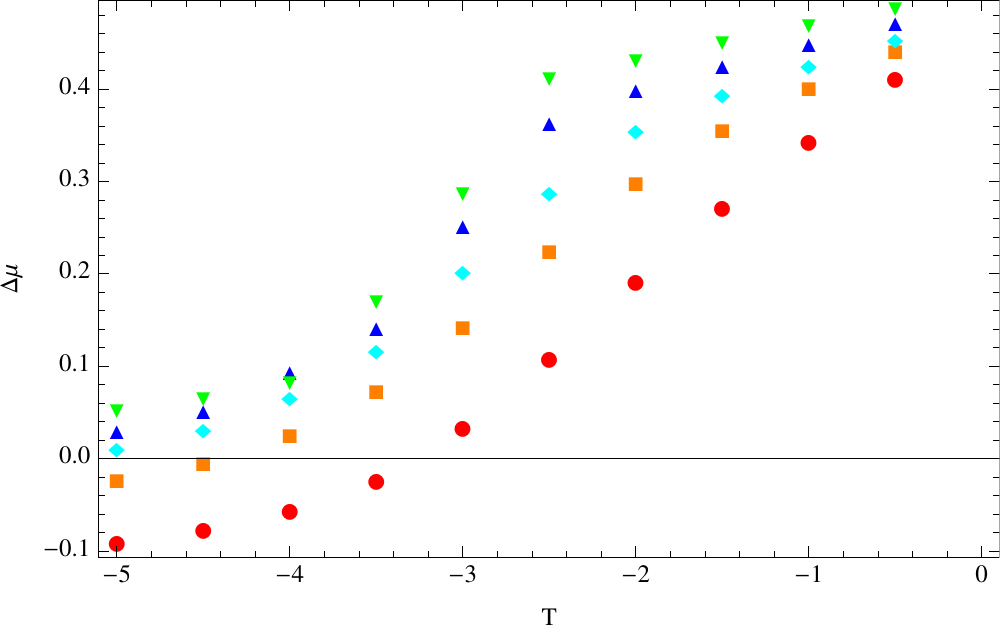}
       \end{center}
\caption{Plots of the relative effective chemical potential as a function of average time for  quenches with $Q =1$  for different conformal dimensions $\Delta = 1.5$  (red circles), $\Delta = 1.35$ (orange squares), $\Delta = 1.25$ (cyan diamonds), $\Delta = 1.15$  (blue triangles) and  $\Delta = 1.01$  (green inverted triangles). }  \label{fig:Deltachempot}
\end{figure}

In Fig.~\ref{fig:chempot} 
and Fig.~\ref{fig:Deltachempot} 
we show how this effective chemical potential $\mu_{eff}(T) $  evolves as a  function  of time for various choices of the net charge density and fermion scaling dimension.  At late times, the system seems to settle in an equilibrium (quasi) Fermi liquid with a Fermi surface that obeys the Luttinger Volume theorem (which states that  the volume enclosed by the Fermi surface is proportional to the fermion charge density) at a finite temperature, chosen to be small compared to the chemical potential in the examples we show.  
In Fig.~\ref{fig:chempot} we examine this effective chemical potential as a function of time for quenches with different values of $Q$ and find several distinct features. As might be anticipated, at $T=0$ the effective chemical potential appears to attain half its final value.  For $T>0$ there is a universality in the approach to the final value of chemical potential, while for $T<0$ the onset of chemical potential gain occurs later for larger $Q$.  
For charges $Q$ of order 1, i.e., comparable to the extremal value $\sqrt3$, the unit of time in  Fig.~\ref{fig:chempot} is of order the inverse final chemical potential. 
Fig.~\ref{fig:chempot} shows that the effective chemical potential is built up on a time scale of the same order.
In Fig.~\ref{fig:Deltachempot} we examine this effective chemical potential for different values of the fermionic operator conformal dimension (bulk mass) and find a final clear feature: the smaller $\Delta$, the quicker chemical potential is acquired.

The effective chemical potential can be viewed as a new non-local probe of thermalization. Probes considered in previous work include equal-time two-point functions, Wilson loops and entanglement entropy \cite{Balasubramanian:2010ce,Galante:2012pv,Caceres:2012em}, in which case the thermalization time depends on the spatial extent of the probe. If one chose probes of size set by the final chemical potential, one would find thermalization times of the same order as in Fig.~\ref{fig:chempot}.


\section{Caveats and further discussion}

In interpreting our results, several caveats need to be taken into account. A first caveat, which we have already mentioned, is that our ``quasi" Fermi liquid realised at long times is in fact a false vacuum artefact associated with the large $N$ limit. 
As we already discussed, it is understood that the true equilibrium state is a Fermi liquid which is dual to the electron star in the bulk. The dynamical ``uncollapse'' of the black brane in the electron star requires a quantised description of the geometry: in classical gravity this cannot happen.  However, as long as $N$ is not too small we expect a separation of time scales such that the further relaxation from the RN ``quasi'' Fermi liquid to the electron star Fermi liquid will take place at a much later time. We leave this problem of the dynamical formation of the electron star due to quantised geometry as a challenge for holography in general.

Next, while our notions of time-dependent spectral function and of time-dependent effective Fermi energy are well-defined and reduce to the appropriate equilibrium concepts in static situations, we should ask to what extent these quantities are measurable in a lab, and exactly which physical information they carry.

In equilibrium, spectral functions can be experimentally determined using Angle-Resolved Photoemission Spectroscopy (ARPES). Actually, electrons can only be emitted from occupied states, so what ARPES measures is really the product of the spectral function and the Fermi factor. Technically, this product equals the ``lesser Green function'' in Fourier space, often denoted $G^<(\omega, k)$ or $D^<(\omega, k)$; see, for instance, \cite{Bellac}. Because of the fluctuation-dissipation theorem, knowledge of the spectral function suffices to determine the lesser Green function, and vice versa (at least for nonzero temperature).

Away from equilibrium, time-resolved ARPES is the tool of choice. As shown in \cite{Freericks}, the lesser Green function (which now depends on two moments of time) still carries the relevant information. In quite close analogy to \eqref{eq:Wignertrans}, it should be Fourier transformed with respect to the relative time coordinate, with the additional complication that the experimental setup introduces time windows for both moments of time. For slowly varying systems, at least some version of the fluctuation-dissipation theorem remains valid \cite{Keller_Jarrell, Mukhopadhyay:2012hv}. Far from equilibrium, however, the retarded and lesser Green functions are not straightforwardly related  (see, for instance, \cite{CaronHuot:2011dr,Chesler:2011ds, Balasubramanian:2012tu}), making it much less clear how ARPES is related to time-dependent spectral functions. While in current versions of time-resolved ARPES the fluctuation-dissipation relation is a good approximation, this would not be the case for an idealized ARPES experiment preformed on our system, which is quenched instantaneously.

One may wonder about the meaning of the oscillations and of the negative regions in our time-dependent spectral functions. Our present understanding is that these are due to simple interference between the pre-quench and post-quench periods, as was the case for the quenched harmonic oscillator discussed in \cite{Balasubramanian:2012tu}. One piece of evidence supporting this statement is that the wavelength of the oscillations in $\omega$ seems to be set by the time $T$ to the quench (see Fig.~\ref{fig:SpectralSection}). This presumably makes it hard to relate this quantity to practical experiments. Given a system prepared in an out-of-equilibrium state, what one would really like to probe is how it behaves {\em after} it has been prepared in that state. But in our setup, the retarded Green function with both moments of time after the quench behaves exactly as if the system were in equilibrium (as discussed in \cite{Bhattacharyya:2009uu} for a thermal quench). So if one defined a time-dependent spectral function using only those ``post-quench'' data (by introducing appropriate time windows), the oscillations and negative regions would disappear. The reason for the equilibrium behaviour is that the post-quench retarded propagator is only sensitive to the bulk geometry outside the shell, which agrees with that of a charged black brane.

Nevertheless, our system right after the quench is far from equilibrium: as also mentioned in \cite{Bhattacharyya:2009uu} (and worked out in detail in \cite{Bhattacharyya:2009uu, Wu:2012rib, Balasubramanian:2010ce, Hubeny:2007xt,AbajoArrastia:2010yt,Albash:2010mv,Liu:2013iza, Balasubramanian:2011at,Allais:2011ys,Callan:2012ip,Hubeny:2013hz}), spatially nonlocal observables do not immediately take their equilibrium values. Therefore, time-resolved ARPES, which does not probe the retarded propagator but the lesser Green function, should not give equilibrium results either (at least in principle: in practice it may not be easy to have good enough time resolution to see deviations from equilibrium). From this point of view (linear response after the quench being trivial, but time-resolved ARPES not), it would be interesting to compute the lesser Green function for our system. While it is in principle known how to do this (see, for instance, \cite{Herzog:2002pc,Skenderis:2008dg}), we leave this computation to future work.


\section*{Acknowledgements}
We thank D.~Dudal, N.~Iqbal,  V.~Ker\"anen and E.~Keski-Vakkuri for useful discussions, M.~Heller for helpful lectures on numerical methods, and Z.~Cao and Y.~Tian for stimulating discussions on the numerical strategy.
B.C.\ and J.Z.\ thank the organizers of the 2014 Amsterdam String Workshop for hospitality during the final stages of this work. This work was supported in part by the Belgian Federal Science Policy Office through the Interuniversity Attraction Pole P7/37, by FWO-Vlaanderen through projects G020714N and G.0651.11, by the Vrije Universiteit Brussel through the Strategic Research Program ``High-Energy Physics'' and by the European Science Foundation Holograv Network. J.Z.\ acknowledges support of a grant from the John Templeton Foundation. The opinions expressed in this publication are those of the authors and do not necessarily reflect the views of the John Templeton Foundation. J.V.\ is Aspirant FWO. D.T.\ and F.G.\ are FWO postdocs.


\begin{appendix}
\section{Numerical convergence}
\begin{figure}[ht]
  \begin{center}
    \includegraphics[width=0.48   \textwidth]{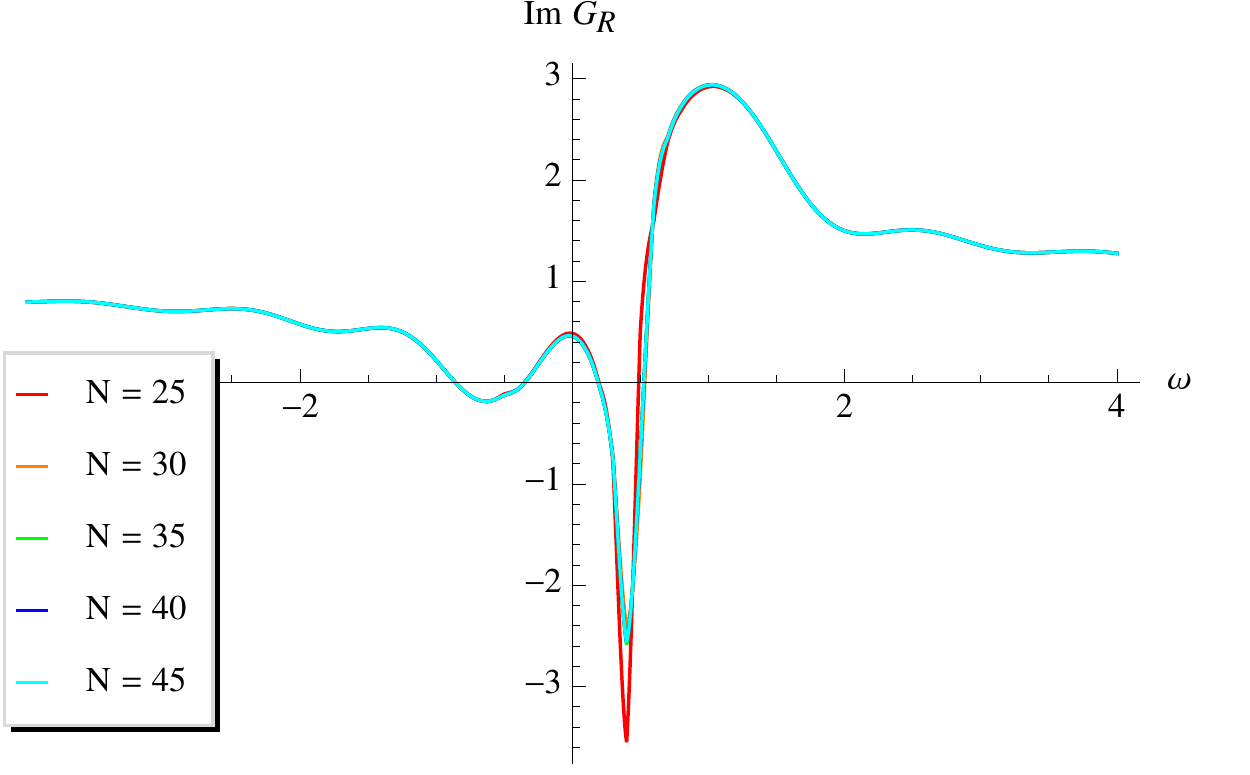}
       \end{center} 
\caption{Convergence as number of Chebyshev nodes is varied  (we fix the Runge-Kutta step  $\delta v=0.05$ and $v_{cut}=60$).  Very good convergence, at the resolution of this plot, is achieved after $N=25$ points. The most sensitive point is the absolute height of the transient negative peak, but even this is well converged at $N=30$.  In the paper we used $N=27$ to balance resource use and accuracy. } \label{fig:convcheb}
\end{figure}

\begin{figure}[ht]
  \begin{center}
    \includegraphics[width=0.48\textwidth]{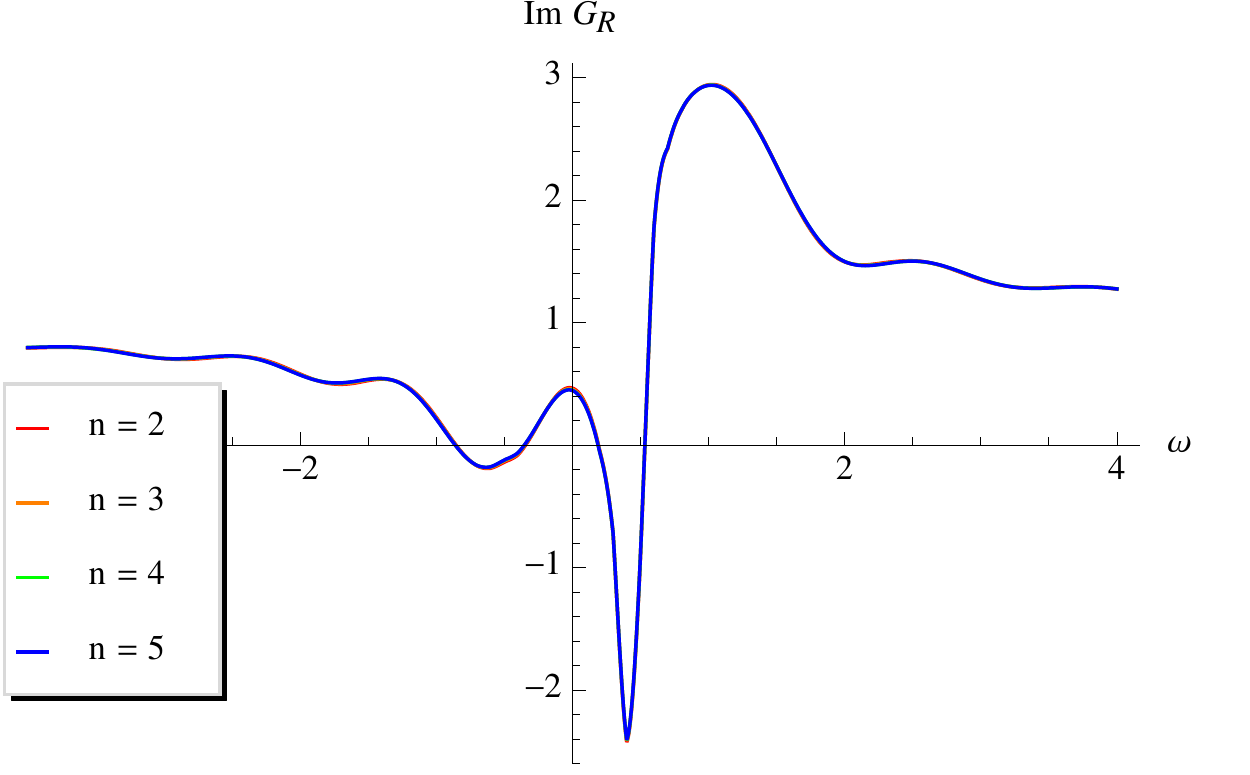}
       \end{center}
\caption{Convergence as Runge-Kutta time step $\delta v= \frac{1}{5}2^{-n}$  is varied (we fix $N=30$ Chebyshev nodes and $v_{cut}=60$).  Very good convergence, at the resolution of this plot the different curves are virtually indistinguishable, is achieved with time stepping $\delta v = 0.05$ as was adopted in the paper. }\label{fig:convRK}
\end{figure}

\begin{figure}[ht]
  \begin{center}
    \includegraphics[width=0.48\textwidth]{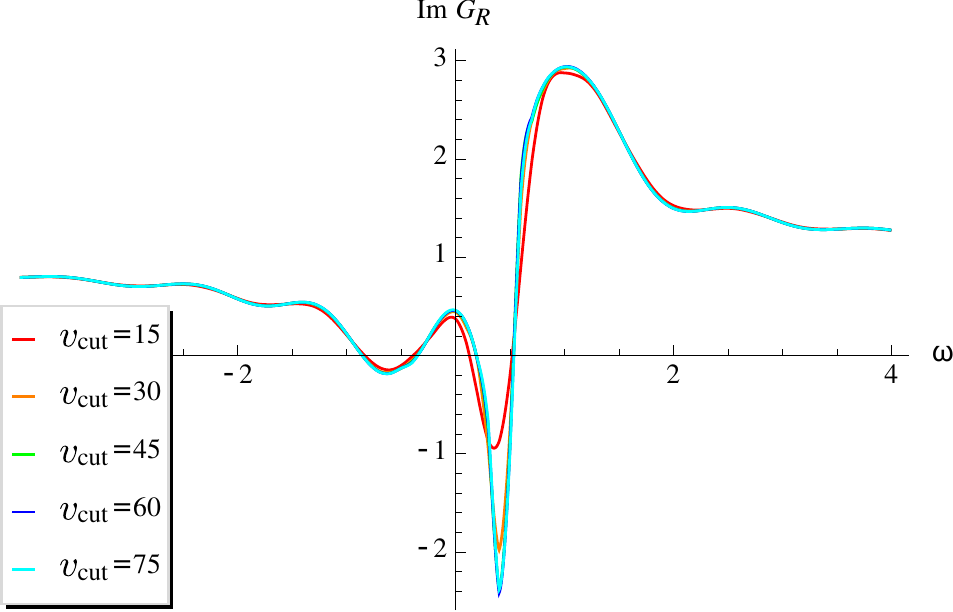}
       \end{center}
\caption{Convergence as the final integration time $v_{cut}$ is varied (we fix $N=30$ Chebyshev nodes and the Runge-Kutta step $\delta v=0.05$).   Here the results are more sensitive but are well converged for $v_{cut} = 45$ (the green, blue and cyan lines  are essentially coincident).  We used $v_{cut}=60$ in the paper. }\label{fig:convcut}
\end{figure}

We provide some details illustrating the convergence of the numerical methods used.  There are three key variables that influence the numerical accuracy: {\it i}) $N$, the number of nodes used in the pseudospectral method to integrate in the radial direction; {\it ii}) $\delta v$, the time-stepping used in the Runge-Kutta integration in the $v$ direction; {\it iii}) $v_{cut}$, the final relative time used for the numerical integration.  The convergence for each  of these is illustrated in Figs.~\ref{fig:convcheb}--\ref{fig:convcut} showing the imaginary part of the Fourier transformed Green's function $G_R(\omega, k)$ for the case of $Q=1.7$, $T=-2.5$ at a fixed $k=0.9$. These values are chosen since they highlight the most challenging parts to obtain numerically. Setting $Q=1.7$, as was used in the bulk of this paper, means that we are at  low temperature and hence $G_{R}(t,k)$ is not thermally suppressed.  The choice  $k=0.9$ corresponds to focusing on the region of momenta space of most interest, where we might anticipate long lived excitations. With $T=-2.5$  we capture the difficult rapidly fluctuating parts of the spectral function arising from transient behaviour in the quench regime. 
\end{appendix}



\end{document}